\documentclass[10pt, twocolumn]{article}

\usepackage{graphicx}%
\usepackage{multirow}%
\usepackage{amsmath,amssymb}
\usepackage{amsthm}%
\usepackage{mathrsfs}%
\usepackage[title]{appendix}%
\usepackage{xcolor}%
\usepackage{textcomp}%
\usepackage{manyfoot}%
\usepackage{booktabs}%
\usepackage{algorithm}%
\usepackage{algorithmicx}%
\usepackage{algpseudocode}%
\usepackage{listings}%
\usepackage{subcaption}
\usepackage{here}
\usepackage{authblk}
\usepackage{url}
\usepackage[T1]{fontenc}
\usepackage{abstract}
\usepackage{here}
\usepackage[top=25truemm, bottom=20truemm,left=14truemm,right=14truemm]{geometry}

\def\tabref#1{Table \ref{#1}}
\def\figref#1{Figure \ref{#1}}
\def\refdatee#1{(accessed #1)}

\usepackage[varg]{txfonts}
\makeatletter%
\input{ot1txtt.fd}
\makeatother%

\def\urle#1{\url{#1}}

\newcommand{\unnumberedfootnote}[1]{%
  \begingroup
  \renewcommand{\thefootnote}{}
  \footnotetext{#1}%
  \endgroup
}

\makeatletter
\renewcommand{\AB@affilsep}{\quad\protect\Affilfont}
\let\AB@affilsepx\AB@affilsep 

\makeatother

\renewcommand{\Affilfont}{\small\itshape}

\title{Interpolation of mountain weather forecasts\\ by machine learning} 
\author{Kazuma Iwase}
\author{Tomoyuki Takenawa}
\affil{Graduate School of Marine Science and Technology, \\
Tokyo University of Marine Science and Technology,\\
2-1-6 Etchujima, Koto-ku Tokyo, 135-8533, Japan}

\date{}

\begin{document}

\twocolumn[
\maketitle
\begin{abstract}
Recent advances in numerical simulation methods based on physical models and their combination with machine learning have improved the accuracy of weather forecasts. However, the accuracy decreases in complex terrains such as mountainous regions because these methods usually use grids of several kilometers square and simple machine learning models. While deep learning has also made significant progress in recent years, its direct application is difficult to utilize the physical knowledge used in the simulation. This paper proposes a method that uses machine learning to interpolate future weather in mountainous regions using forecast data from surrounding plains and past observed data to improve weather forecasts in mountainous regions. We focus on mountainous regions in Japan and predict temperature and precipitation mainly using LightGBM as a machine learning model. Despite the use of a small dataset, through feature engineering and model tuning, our method partially achieves improvements in the RMSE with significantly less training time.\\

\vskip 5mm
\end{abstract}
]

\section{Introduction}
\unnumberedfootnote{E-mail address: kazumaiwase676@gmail.com, takenawa@kaiyodai.ac.jp}
The rising interest in leisure activities, drone transportation to remote areas, and enhanced on-site operations efficiency have increased the demand for accurate and comprehensive weather forecasting in mountainous regions. Recent advancements in numerical simulation methods known as Numerical Weather Prediction (NWP), such as Global Spectral Model (GSM) and Meso-Scale Model (MSM) of Japan Meteorological Agency (JMA) \footnote{\url{https://www.data.jma.go.jp/risk/obsdl/index.php}}, have improved weather forecast accuracy (Chapter 3 of \cite{outline2023}). However, the accuracy decreases in complex terrains like mountainous regions due to the several kilometers square grid used in numerical simulations \cite{goger2016current, golzio2021land}. To reduce systematic errors in NWP outputs, JMA provides various types of forecast guidance, such as MSM Guidance (MSMG) (Chapter 4 of \cite{outline2023}), which applies machine learning models to the output of MSM. Although MSMG still deals with 5 kilometer square grids for precipitation forecasts, AMeDAS \footnote{\url{https://www.jma.go.jp/jma/en/Activities/amedas/amedas.html}} points are subject to time series temperature forecasts of MSMG. While deep learning has also significantly advanced, its direct application is difficult to utilize physics knowledge used in the simulation.

In this paper, we propose a method that employs machine learning to ``interpolate'' future weather in mountainous regions using current observed data and forecast data from surrounding plains. Specifically, we predict the temperature and precipitation at Mt. Fuji and Hakone respectively, which are located in mountainous areas of the Kanto region in Japan, which is characterized by a warm and humid climate. These predictions are made for 2, 7, 8, and 9 hours ahead. We then compare the accuracy of various machine learning models and existing weather forecast services.

Compared to direct use of NWP data or guidance data, this approach is more accessible to general users as it utilizes past observed data and forecast data from weather services, which are readily available. In particular, our method uses only historical observed data during training, which makes data collection easy.

Furthermore, we examine the effectiveness of using a linear combination of the Mean Squared Error (MSE) and binary cross-entropy as a loss function, given that precipitation values are non-negative and the occurrence of rainfall is important.

Despite the use of a small dataset, through feature engineering and model tuning, our method partially achieves improvements in the Root Mean Squared Error (RMSE) with significantly less training time.

\section{Related works}
Machine learning models used in weather forecasting can be divided into two categories: time series models that take a sequence of data with a specific temporal structure as input, and general regression models that do not require such structured input data. In time series models, it is common to use Recurrent Neural Networks (RNN), Long Short-Term Memory (LSTM), Convolutional Neural Networks (CNN), and their variants \cite{mehrkanoon2019deep, shi2015convolutional, zaytar2016sequence} (or see references in \cite{bilgin2021tent}). However, recently, a time series model based on Self-Attention called Tensorized Encoder Transformer (TENT) has been proposed, which has been reported to outperform RNN and LSTM in terms of accuracy \cite{bilgin2021tent}. On the other hand, As regression models, Linear Regression, Polynomial Regression, Lasso Regression, Ridge Regression, Support Vector Regression and Random Forest Regression are evaluated for temperature predictions \cite{jahnavi2019analysis}. ARIMA models and conventional Neural Networks (Multilayer Perceptrons) are used in studies such as \cite{chen2011comparison} and \cite{kuligowski1998localized}. 
JMA uses several simple machine learning methods such as Linear Regression, Kalman Filtering and Neural Networks to refine prediction (Chapter 4 of \cite{outline2023}).
Similar to this paper, \cite{yoshikane2022bias} uses Support Vector Machine (SVM) to interpolate numerical simulation results from NWP.

Regarding precipitation prediction, the study proposed in \cite{larraondo2020optimization} uses binary categorical loss functions, employing an original loss function based on binary classification problem metrics.
However, in this paper, we use a linear combination of the binary cross-entropy and the MSE as a loss function.

\section{Our Method}
In this paper, we utilize not only the observed data up to the present but also future forecast data from the surrounding areas of the target location. While using future data might seem like data leakage, it is not an issue as the forecast data from neighboring regions is available at prediction time. However, we do not use forecast data for mountainous regions, including the target location, as we assume that forecast data for mountainous regions have low reliability \footnote{In this study, we use ``forecasts" to refer to estimations of weather services and ``predict" or ``predictions" to describe our estimations.}.

Due to the formal and qualitative differences in the data before and after the current point, this paper does not employ time series models.  Instead, it employs regression models such as Linear model (Elastic Net) \cite{zou2005regularization}, LightGBM \cite{ke2017lightgbm}, XGBoost \cite{chen2016xgboost}, Random Forest (Extra Trees) \cite{geurts2006extremely}, and Neural Networks (NN, Multi Perceptron).
Here, Elastic Net is a linear model with L1 and L2 regularization, and thus, learning can be done in a short time, but accuracy is limited due to the lack of expressive ability. Random Forest is a simple ensemble method of decision trees.
XGBoost is also an ensemble method of decision trees but adapts the boosting algorithm
that sequentially adds decision trees and a quadratic optimization method. LightGBM is a lighter version of XGBoost. XGBoost and LightGBM are considered to have a good balance of bias and variance and are known to perform particularly well for table data \cite{shwartz2022tabular}. NN are general-purpose models, but they take time to train, and have generalization performance problems for table data.
Since these regression models except NN are not expected to automatically extract features, and since some improvement in accuracy can be expected for NN as well, lag features and moving averages, which are variables commonly used in time series analysis, are used as input features.

Furthermore, the regression models other than NN are faster than deep learning-based time series models, enabling efficient training, inference, feature selection, and tuning in shorter time.

\begin{figure}[t]
\centering
\includegraphics[keepaspectratio, width=1\hsize]{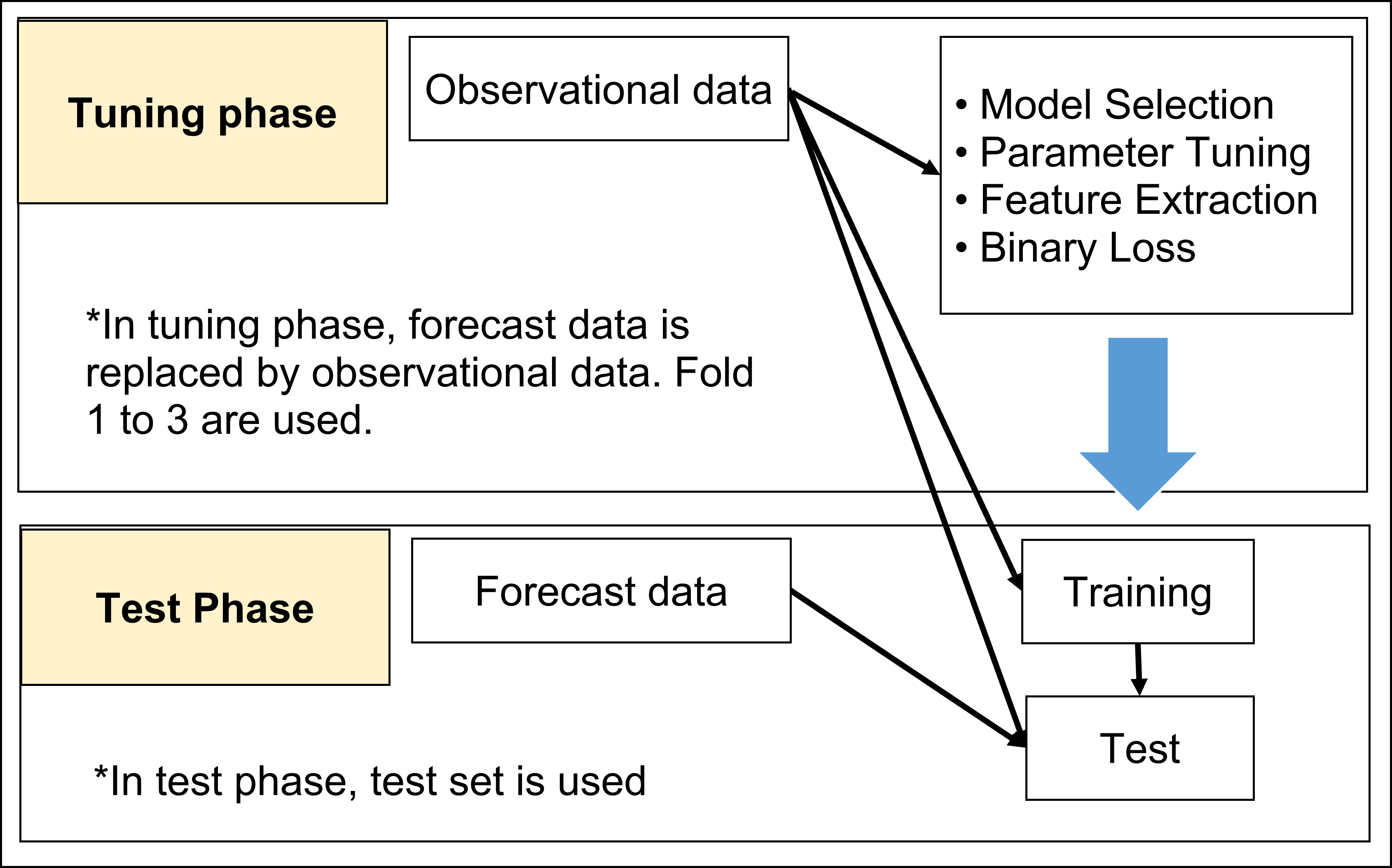}
\caption{Overview of the tuning and test phase.}
\label{fig:overview}
\end{figure}

Due to the difficulty in obtaining long-term forecast data, we only use observed data for model performance comparisons, model tuning, and feature selection, and use forecast data only for evaluating the general performance of LightGBM, which outperforms other models for observed data (\figref{fig:overview}).

We mainly use the MSE for the loss function and the RMSE for the evaluation metric:
\begin{equation}
\label{eqrmse}
{\rm RMSE} = \sqrt{\frac{1}{N} \sum_{i=1}^{N} (y_i - \hat{y})^2}
\end{equation}
where $N$, $y$ and $\hat{y}$ are the number of samples in the dataset,
the target value (ground truth)  and predicted value respectively.
However, the observed precipitation data used in this experiment takes only non-negative values with a resolution of 0.5 [mm] and it is important to determine whether the precipitation is exactly 0 [mm], 0.5 [mm] or higher. To incorporate this into training, we use a linear combination of the MSE and the shifted binary cross entropy 
\begin{equation} \label{binaryloss}
L = \alpha {\rm MSE}(y, \hat{y}) + (1-\alpha) L_{{\rm binary}}, 
\end{equation}
as a loss function in some experiments, where $\alpha \geq 0$ is a constant.
See  Appendix A.1 for details on this loss function.  

We also use Correlation Coefficient (CC) as another evaluation metric 
\begin{equation} \label{correlation coefficient}
{\rm CC} = \frac{\sum_{i=1}^{n}(y_i - \bar{y})(\hat{y}_i - \bar{\hat{y}})}{\sqrt{\sum_{i=1}^{n}(y_i - \bar{y})^2 \sum_{i=1}^{n}(\hat{y}_i - \bar{\hat{y}})^2}},
\end{equation}
where $\bar{y}$ and $\bar{\hat{y}}$ are the mean of the target values and predicted values respectively.

\section{Experiments}
The code of this study is available on GitHub page \footnote{\url{https://github.com/KazumaIwase/interpolation_ml}} where the forecast data from ``Weathernews'' or ``Tenki to Kurasu'' has been replaced with artificially created dummy data.

\subsection{Data}
\begin{figure}[t]
\centering
\includegraphics[keepaspectratio, width=0.9\hsize]{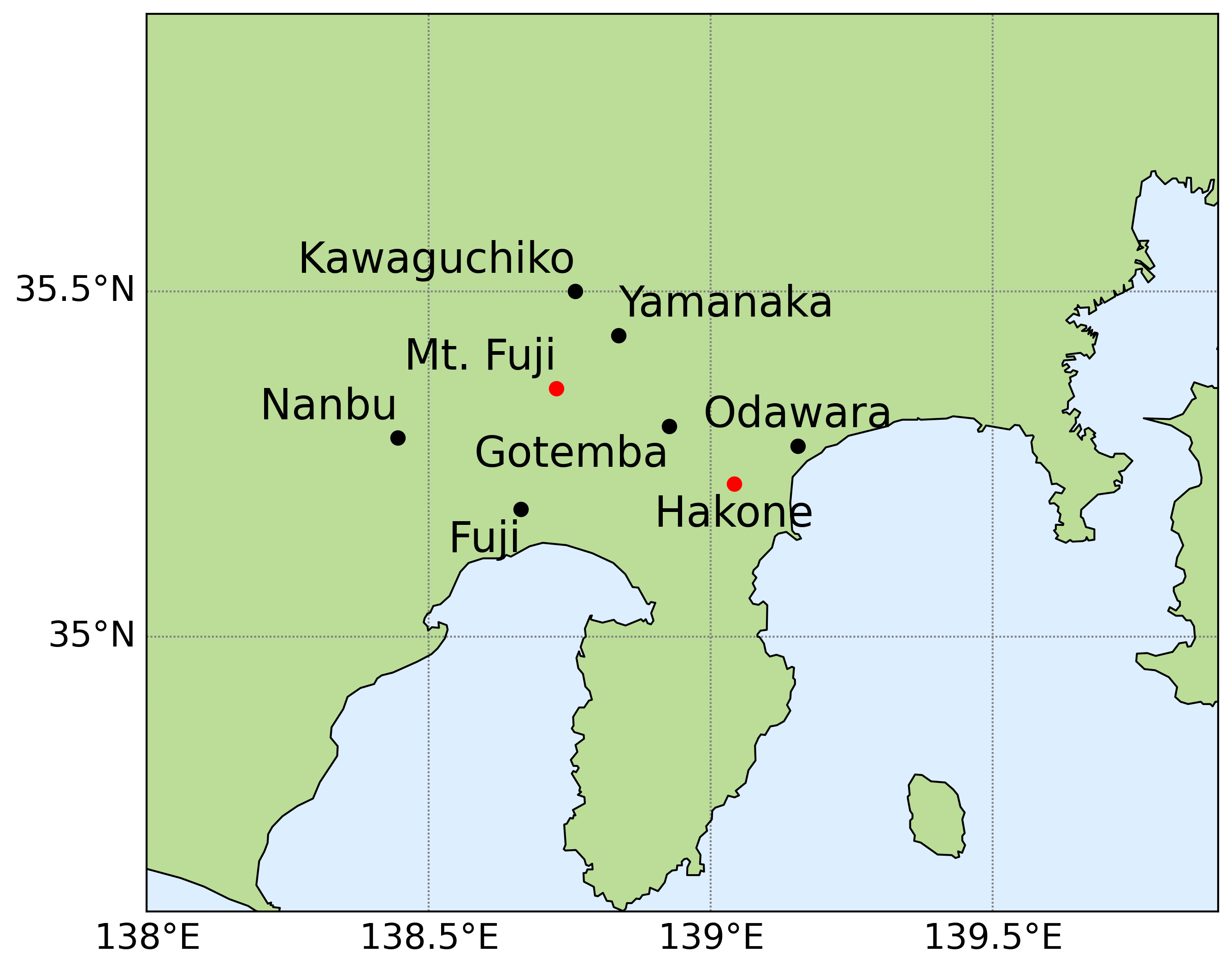}
\caption{Observation points. Black points represent surrounding areas and red ones represent target mountainous areas.}
\label{fig:map}
\end{figure}

The experiment aims to predict temperature at the summit of Mt. Fuji (Fujisan) (elevation 3,775 m) and precipitation at Hakone (elevation 855 m), which are among a few mountain meteorological data that are consistently provided by the JMA. Mt. Fuji is the highest peak in Japan, and meteorological observations including temperature, pressure, and humidity are recorded by the agency. Hakone is a complex volcano located approximately 30-40km southeast of Mt. Fuji (\figref{fig:map}), with the interior of the outer rim forming a highland. The observation point of the agency is located at the foot of the central crater cone within the outer rim, where only precipitation is measured.

The surrounding areas include Odawara (temperature, precipitation, wind speed), Kawaguchiko (temperature, wind speed), Gotemba (temperature, precipitation, wind speed), Yamanakako (temperature, precipitation, wind speed), Nambu (temperature, precipitation, wind speed), and Fuji (temperature, precipitation, wind speed) (\figref{fig:map}).

\begin{table}[t]
\centering
\caption{RMSE of forecast services and observed data. Upper columns represent surrounding points and lower columns represent mountain points. Tenki (3775m) is the weighted average of ``Tenki to Kurasu'' from 3100m and from 4400m, the others are from ``Weathernews''.}
\label{tab:RMSETest}
\footnotesize
\begin{tabular}{c|cc|cc}\hline\hline
& \multicolumn{2}{c|}{Temperature} &
\multicolumn{2}{c}{Precipitation}  	\\ 
&2 hours &8 hours &2 hours &8 hours \\ \hline
Fuji&0.867 &1.595 &1.033 &0.614  \\ 
Gotenba&1.247 &1.294 &0.685 & 0.303 \\ 
Odawara&1.363 &1.387 &0.393 &0.315  \\ 
Yamanaka&1.467 &1.193 &0.973 &0.349  \\ \hline
Mt. Fuji&7.133 &4.852 &- &-  \\ 
Hakone& -&- &1.457 &0.443 \\
Tenki (3775m)&4.296&3.627&&\\ 
\end{tabular}
\end{table}

\begin{table*}[t]
\centering
\caption{Features used in this study. Surround and Target represent the surrounding areas and the target locations respectively. Observe(2h ago) represents an observed value from 2 hours ago. Forecast(1h ago) represents a forecast value from 1 hour ago. Diff(2,3h ago) represents a difference between values from 2, 3 hours ago and their respective values from 1 hour ago. Diff((8h,7d ago)) represents a difference between a value from 8 hours ago and a value from 7 days ago starting 8 hours ago. Avg, Max, Min(2h-24h) represents average, maximum, and minimum observed values of past 24 hours starting 2 hours ago. Features at 7 and 9 hour prediction are almost the same as the 8 hour prediction. Forecast data is replaced with observed data during training.}
\label{tab:features}
\footnotesize
\begin{tabular}{lp{0.15\linewidth}p{0.21\linewidth}p{0.18\linewidth}p{0.18\linewidth}}\hline\hline
& \multicolumn{2}{l}{Mt. Fuji} & \multicolumn{2}{l}{Hakone}  	\\ 
&2 hour &8 hour &2 hour &8 hour \\ \hline
Seasonality& \multicolumn{2}{l}{Time of a day and Day of a year represented as sin, cos functions} & \multicolumn{2}{l}{Day of a year represented as sin, cos functions}  	\\ \hline
\multirow{4}{*}{Surround and Target}&Observe(2,3,4,5h ago),\par\noindent Diff(2,3h ago) &Observe(8,9,10,23,24,25h ago),\par\noindent Diff(8,9,24h ago),\par\noindent Diff((8h,24h ago),(8h,7d ago)),\par\noindent Diff((24h,24h ago),(24h,7d ago))  &Observe(2,3,4,5h ago),\par\noindent Diff(2,3h ago) &\multirow{4}{*}{-}  \\ \hline
\multirow{2}{*}{Only Surround}&Forecast(0,1h ago),\par\noindent Diff(0h ago) &Forecast(0,1,2,3,4h ago),\par\noindent Diff(0,1,2h ago)  &Forecast(0,1h ago),\par\noindent Diff(0h ago)  &Forecast(0,1,2,3,4,5h ago),\par\noindent Diff(0,1,2,3,4h ago)   \\ \hline
\multirow{2}{*}{Only Target}&Avg, Max, Min\par\noindent(2h-24h,2h-7d ago) &Avg, Max, Min\par\noindent(8h-24h,8h-7d ago) &Avg, Max, Min\par\noindent(2h-12h,2h-24h,2h-7d ago)  &Avg, Max, Min\par\noindent(8h-12h,8h-24h,8h-7d ago)  \\ \hline
\end{tabular}
\end{table*}

\begin{table*}[t]
\centering
\caption{
Data folding for cross validation and test data. 
This table is in the case of 2 hour prediction at Mt. Fuji.  
The left numbers represent the number of hourly data and the right parentheses include time periods. Fold 1 to 3 are made for cross validation along the time series. Training data for the test phase consist of entire training and validation time periods. Test data in bold are hourly data taken once a day. For the other predictions, starting periods and the number of hourly data change slightly depending on a prediction time and a target area because different types of feature engineering are performed as shown in Table 2. 
}
\label{tab:Dataset}
\footnotesize
\begin{tabular}{l|ll} \hline\hline                       
Data                  & Train & Validation / \textbf{Test} \\ \hline
Fold 1                 &20057 (2019/07/03 15:00:00-2021/10/16 07:00:00) & 2920 (2021/10/16 08:00:00-2022/02/14 23:00:00)                           \\
Fold 2                 & 22977 (2019/07/03 15:00:00-2022/02/14 23:00:00) & 2920 (2022/02/15 00:00:00-2022/06/16 15:00:00)                         \\
Fold 3                 & 25897 (2019/07/03 15:00:00-2022/06/16 15:00:00) & 2920 (2022/06/16 16:00:00-2022/10/16 07:00:00)                        \\ \hline
Test Set           & 28817 (2019/07/03 15:00:00-2022/10/16 07:00:00) & \textbf{{\phantom 0}144 (2022/10/16-2023/03/08)}
\end{tabular}
\end{table*}

Observed data were collected from the JMA, consisting of hourly observations for all locations and variables mentioned above from June 26, 2019 14:00:00 to March 9, 2023 00:00:00 in Japan time. The hourly forecast data for the surrounding areas were obtained by web scraping from ``Weathernews'' \footnote{\url{https://Weathernews.jp/}} at 7 a.m. from October 16, 2022 to March 8, 2023, covering 144 days. Furthermore, to validate the prediction accuracy, hourly forecast data for the temperature at Mt. Fuji and precipitation at Hakone from Weathernews were collected at the same time during the same period as the surrounding forecasts, along with hourly forecast data for the elevations of 3,100 m and 4,400 m at Mt. Fuji from ``Tenki to Kurasu'' \footnote{\url{https://tenkura.n-kishou.co.jp/tk/}}. Compared to grid data such as MSMG, these data are considerably small as we only use data from 8 points (\figref{fig:map}).

\tabref{tab:RMSETest} shows  the RMSE of forecast services. It is evident that the temperature at Mt. Fuji has a significantly larger RMSE compared to the surrounding areas and the precipitation at Hakone has a slightly larger RMSE, except for 8h forecasts of Fuji.

In this experiment, during the test phase, forecast data for the surrounding area is obtained at 7:00 a.m., and then predictions for several hours into the future are made and compared with actual observed and forecast data. However, during the training phase, only observed data is used. In other words, the future forecast data is replaced with observed data for training, as it was not feasible to collect forecast data for the entire training period (approximately 3 years).

Data preprocessing includes missing value imputation for which linear interpolation method is used. Subsequently, feature engineering is performed as in \tabref{tab:features}. In this experiment, when predicting the temperature at Mt. Fuji 8 hours ahead, 8 hours ahead is used as the reference point, i.e. ``8 hours ago'' in \tabref{tab:features} implies the time when the prediction is made.

Basically following the case of the Autoregressive Integrated Moving Average (ARIMA) model, we select the input features from time series data, their differences, and moving averages.
However, the main difference in our case is that we use multiple time series. We do not use the forecast data for the target location since we assume that the forecast has a larger error in mountainous areas. In addition, we do not use differences and moving averages across observed and forecast data since the data varies depending on whether the data is in the past (observed data) or in the future (forecast data).

During the tuning of hyperparameters, the time-series data divided for cross-validation is used, while during testing, the model trained with all the available training data is used (\tabref{tab:Dataset}). It should be noted that although the three validation data combined cover one year, the test data covers a period of 144 days centered around the winter season, which may lead to data bias.

\begin{table}[t]
\centering
\caption{RMSE and training times (iterations) of the models. The upper and lower numbers are the averages of the RMSE on the three validation data in Fold 1 to 3, and training times on the training data in test set respectively. The unit of training time is seconds. Iterations are gained by early stopping on Fold 3.}
\label{tab:models}
\footnotesize
\begin{tabular}{c|cc|cc} \hline\hline
& \multicolumn{2}{c|}{\footnotesize
\begin{tabular}[c|]{@{}c@{}}Temperature \\ at Mt. Fuji \end{tabular}} &
\multicolumn{2}{c}{\footnotesize
\begin{tabular}[c]{@{}c@{}}Precipitation \\ at Hakone \end{tabular}}  	\\
&2 hours &8 hours &2 hours &8 hours \\ \hline
\multirow{2}{*}{Elastic Net} &1.178 &2.188 &1.063 &1.093   \\ 
           &3.2	 & 6.2 &\textbf{0.1}  & \textbf{0.1} \\ \hline
\multirow{2}{*}{LightGBM} &\textbf{1.109} &  \textbf{2.006}  &\textbf{0.951}&  0.999 \\ 
           &3.2 (124) &  8.2 (174) & 0.8 (27) & 0.9 (16) \\  \hline
\multirow{2}{*}{XGBoost} &1.133 &   2.069  &0.971&\textbf{0.995}   \\ 
           &\textbf{1.1} (43) &  \textbf{3.0} (39) &0.8 (24) &  0.7 (6) \\ \hline
\multirow{2}{*}{Extra Trees} &1.140 & 2.101 &1.024  &  1.085  \\ 
           &79.1 &  170.8 &76.0 &81.5    \\ \hline
\multirow{2}{*}{Neural Net} & 1.260&  2.187 & 1.014  &  1.086    \\ 
           &84.6 (74) &  36.3 (31) &49.0 (42)  &  45.7 (39)
\end{tabular}
\end{table}

\begin{table}[t]
\centering
\caption{RMSE losses of precipitation prediction at Hakone. The values are calculated as the average of the RMSE of the three validation data.}
\label{tab:binary_val}
\footnotesize
\begin{tabular}{l|cccc} \hline\hline
Model     & 2 hours        & 7 hours & 8 hours & 9 hours \\ \hline
LGBM (all)    &   \textbf{0.946}     & 0.994     & 0.998        &  \textbf{0.997}      \\
LGBM (top)     &      0.958          & 0.999      & \textbf{0.987}   & 0.998 \\
LGBM binary (all)    &   0.965  & 0.996     & 0.998      &  1.002       \\
LGBM binary (top)  &   0.960         & \textbf{0.990}       & 1.001  &  1.001   \\
\end{tabular}
\end{table}

\begin{figure*}[t]
\centering
\begin{minipage}{0.48\hsize}
\centering
\subcaption{Temperature in 2 hours at Mt. Fuji}
\includegraphics[keepaspectratio, width=1.0\hsize]{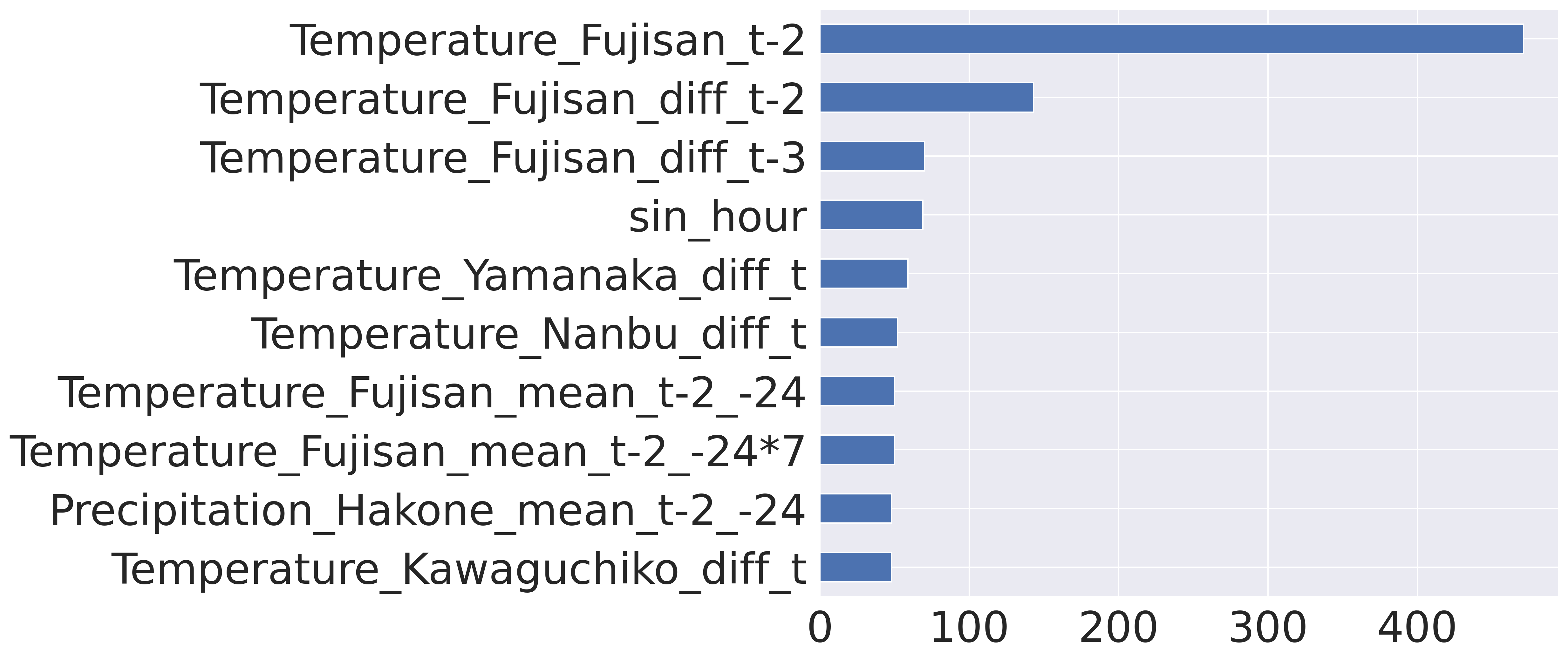}      
\end{minipage} 
\begin{minipage}{0.48\hsize}
\centering
\subcaption{Precipitation in 2 hours at Hakone}
\includegraphics[keepaspectratio, width=1.0\hsize]{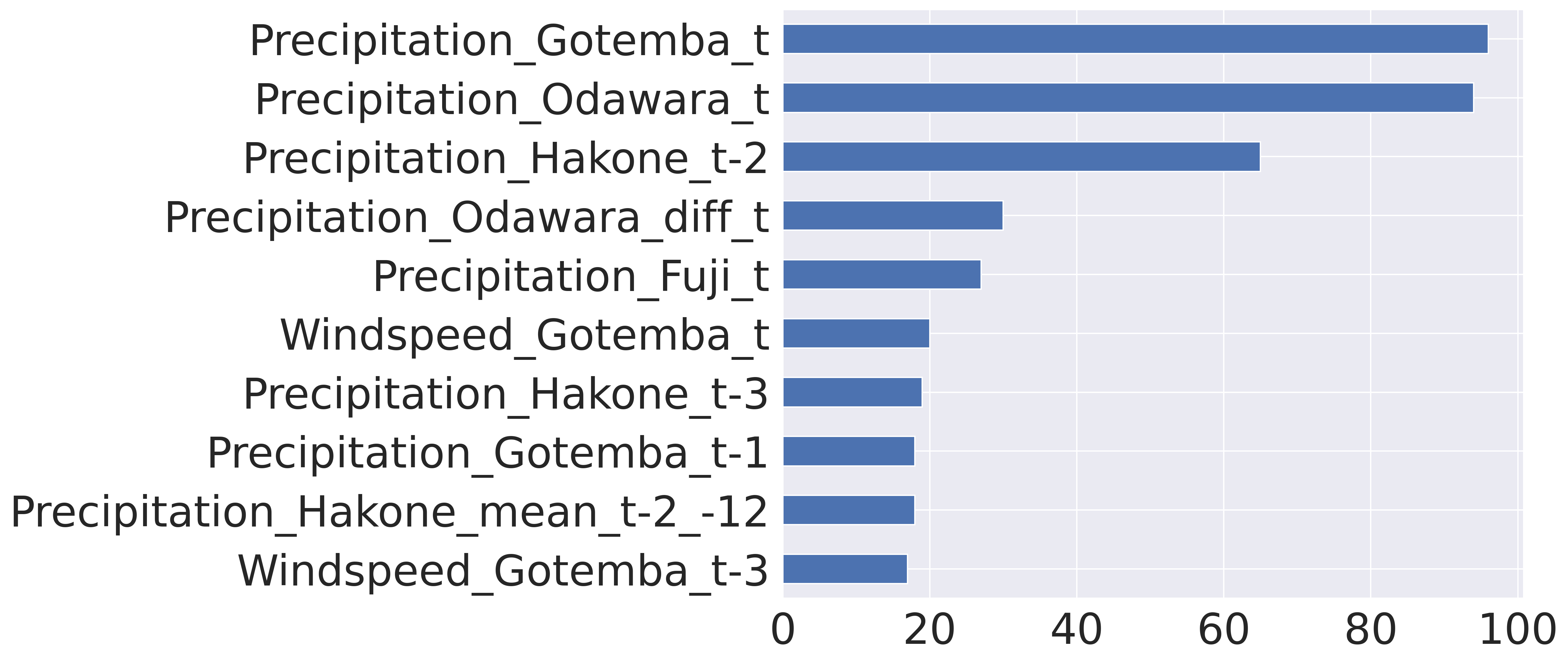}
\end{minipage}\\
\begin{minipage}{0.48\hsize}
\centering
\subcaption{Temperature in 8 hours at Mt. Fuji}
\includegraphics[keepaspectratio, width=1.0\hsize]{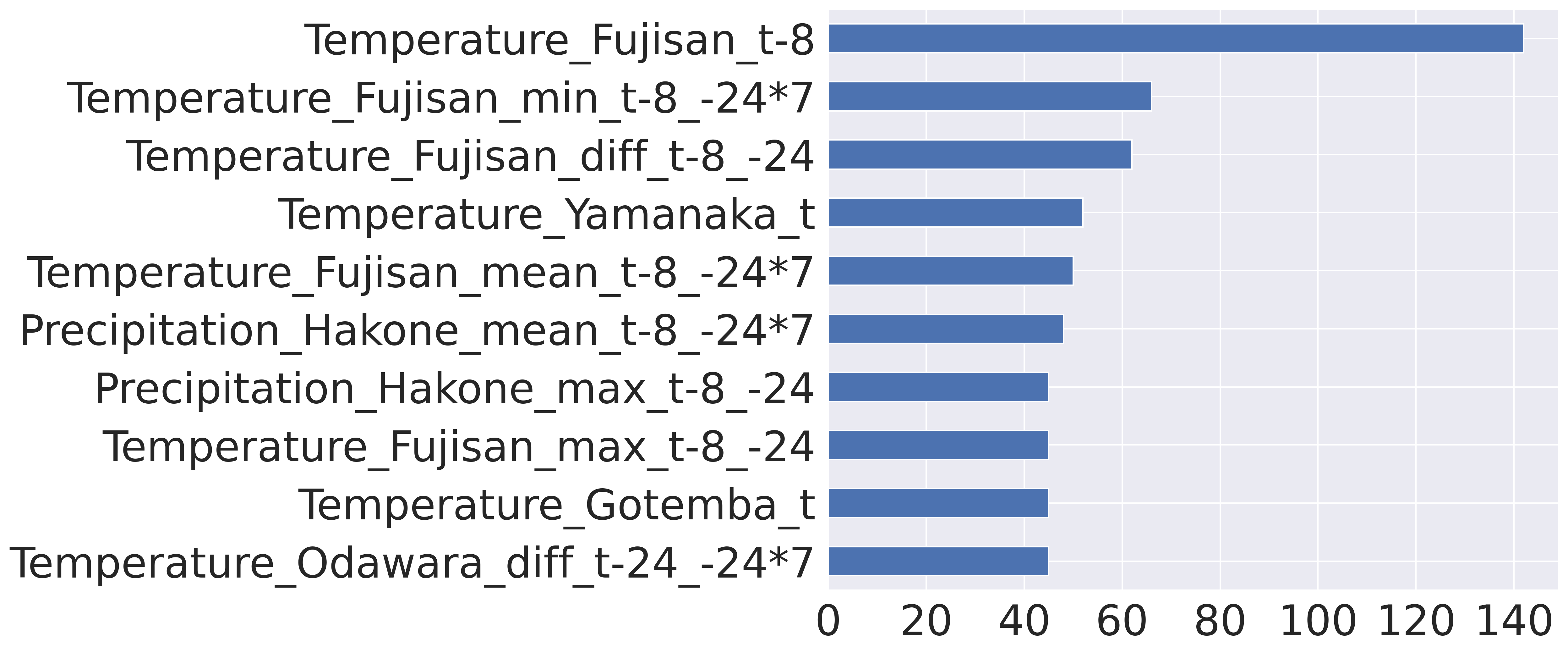}
\end{minipage} 
\begin{minipage}{0.48\hsize}
\centering
\subcaption{Precipitation in 8 hours at Hakone}
\includegraphics[keepaspectratio, width=1.0\hsize]{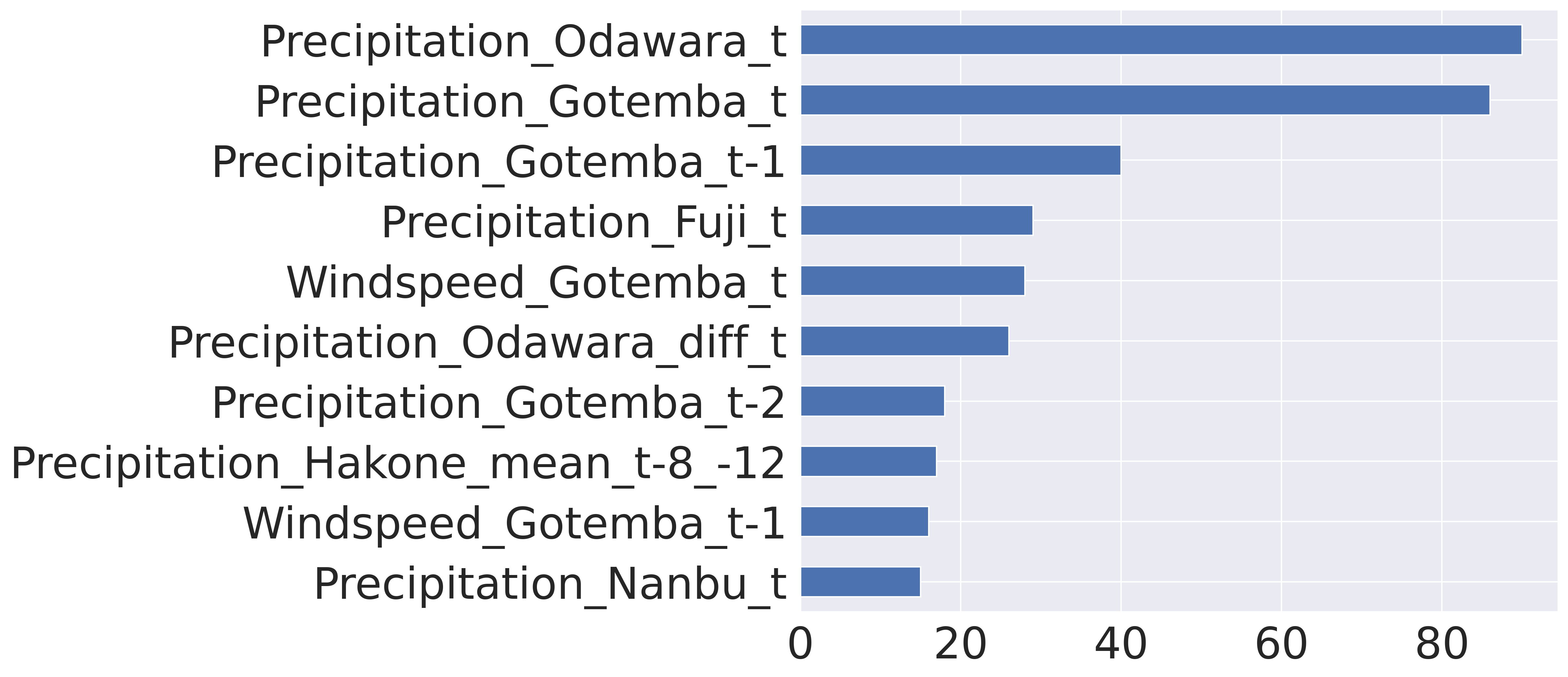}
\end{minipage}\\
\caption{Top 10 important features of LightGBM. LGBM (all) and LGBM binary (all) are used for prediction of temperature at Mt. Fuji and precipitation at Hakone, respectively, where $t\mbox{-}a$ means start at $a$ hours ago from the target time, while $t\mbox{-}a\_b$ means start at $a$ hours ago and end at $a+b$ hours ago.}
\label{fig:importance}
\end{figure*}

\begin{table*}[t]
\centering
\caption{RMSE losses and training times (iterations) of predictions/forecasts on the test data. We use a threshold procedure for precipitation that makes the predicted value zero if it is less than 0.3 [mm], but its effect is very small like $\pm 0.01$ for the RMSE loss. The unit of training time is seconds. Iterations are gained by early stopping on Fold 3.}
\label{tab:rmse}
\footnotesize
\begin{tabular}{l|l|cccc|cccc} \hline\hline
\multirow{2}{*}{Feature at Station} & \multirow{2}{*}{Model and Forecast}  & \multicolumn{4}{c|}{\footnotesize
\begin{tabular}[c|]{@{}c@{}} RMSE \end{tabular}} &
\multicolumn{4}{c}{\footnotesize
\begin{tabular}[c]{@{}c@{}} Training Time (Iteration) \end{tabular}} \\ & & 2 hours & 7 hours & 8 hours & 9 hours  & 2 hours & 7 hours & 8 hours & 9 hours \\ \hline
\multirow{6}{*}{Temperature at Mt. Fuji}   
& LGBM (all)   &  \textbf{1.068} & \textbf{1.944} & \textbf{2.015} & 2.242 & 3.7 (123) & 7.9 (187)  & 7.7 (137) & 7.2 (101)    \\
& LGBM (top)  & 1.069 & \textbf{1.944}   &  2.097  &  \textbf{2.216}& 4.2 (116) & 2.9 (152)   &  2.7 (103)  &  2.8 (140)   \\ 
& LGBM (no future + all)  & 1.112 & 2.144   &  2.286  &  2.488& 2.1 (85) & 3.1 (152)   &  4.5 (103)  &  6.0 (172)  \\ 
& LGBM (no future + top)    &  1.099  &  2.122   &  2.310  &2.449& 1.8 (73) & 1.5 (148)   &  2.0 (126)  &  2.5 (137) \\ \cline{2-10}
& Weathernews     & 7.133   & 5.279 & 4.852   & 3.902 & - & -   &  -  &  -  \\
& Tenki (3775m)   & 4.296          & -     & 3.627   & - & - & -   &  -  &  -   \\ \hline
                                     
\multirow{9}{*}{Precipitation at Hakone}
& LGBM (all)    &   1.046             & 0.486     & 0.454        &  0.623 & 1.5 (28) & 1.2 (17)   &  1.1 (18)  &  1.4 (16)     \\
& LGBM (top) &      1.098         & 0.496       & 0.467        & 0.634 & 0.2 (35) & 0.2 (16)   &  0.2 (19)  &  0.2 (13)\\
& LGBM binary (all)    &   1.108  & 0.457  & \textbf{0.381}   &  0.586 & 1.6 (46) & 1.4 (28)   &  1.5 (28)  &  1.6 (29)    \\
& LGBM binary (top)  &   1.186	         & 0.419    & 0.388   &  \textbf{0.512}  & 0.2 (33) & 0.3 (37)   &  0.3 (36)  &  0.2 (30) \\
& LGBM (no future + all)    &   1.083  & 0.734     & 0.828   &  0.850  & 1.0 (14) & 0.7 (44)   &  0.7 (15) &  0.6 (12)    \\
& LGBM (no future + top)    &   1.028 & 0.749     & 0.868   &  0.783  & 0.1 (16) & 0.1 (21)   &  0.2 (26)  &  0.2 (11)    \\
& LGBM binary (no future + all) & \textbf{0.691} & 0.596 & 0.770  &  0.895  & 1.4 (49) & 0.7 (37)   &  0.9 (37)  &  0.7 (14)    \\
& LGBM binary (no future + top) &  1.096 &  0.636 &  0.779   & 0.895  & 0.2 (22) & 0.1 (5)  & 0.3 (28)  & 0.2 (22) \\ \cline{2-10}
& Weathernews            & 1.457          & \textbf{0.344} & 0.443   & 0.619  & - & -   &  -  &  -   \\         
\end{tabular}
\end{table*}

\begin{table*}[t]
\centering
\caption{Correlation Coefficient of predictions/forecasts on the test data. We use a threshold procedure for precipitation that makes the predicted value zero if it is less than 0.3 [mm]. Nan appeared when all the predicted values were zero.}
\label{tab:cc}
\footnotesize
\begin{tabular}{l|l|cccc} \hline\hline
\multirow{2}{*}{Feature at Station} & \multirow{2}{*}{Model and Forecast}  & \multicolumn{4}{c}{\footnotesize
\begin{tabular}[c]{@{}c@{}} Correlation Coefficient \end{tabular}} \\ & & 2 hours & 7 hours & 8 hours & 9 hours  \\ \hline
\multirow{6}{*}{Temperature at Mt. Fuji}   
& LGBM (all)   &  \textbf{0.987} & 0.953 & \textbf{0.947} & 0.934  \\
& LGBM (top)  & \textbf{0.987} & \textbf{0.954}   &  0.943  &  \textbf{0.935}   \\ 
& LGBM (no future + all)  & \textbf{0.987} & 0.945   &  0.933  &  0.917  \\ 
& LGBM (no future + top)    &  \textbf{0.987}  &  0.946   &  0.931  &0.921 \\ \cline{2-6}
& Weathernews     & 0.486   & 0.868 & 0.873   & 0.888 \\
& Tenki (3775m)   & 0.951          & -     & 0.944 & -  \\ \hline
                                     
\multirow{9}{*}{Precipitation at Hakone}
& LGBM (all)    &   0.661            & 0.783     & 0.843        &  0.722   \\
& LGBM (top) &      0.614         & 0.806	       & 0.833        & 0.701 \\
& LGBM binary (all)    &   0.602  & 0.858  & \textbf{0.898}   &  0.748    \\
& LGBM binary (top)  &   0.519		 &\textbf{0.862}    & 0.880   &  \textbf{0.812} \\
& LGBM (no future + all)    &   0.680  & 0.321     & 0.356   &  0.300     \\
& LGBM (no future + top)    &   0.759 & 0.302     & 0.275   &  0.452    \\
& LGBM binary (no future + all) & \textbf{0.908} & 0.369 & 0.357  &  -0.018   \\
& LGBM binary (no future + top) &  0.827 &  -0.016 &  0.235   & Nan  \\ \cline{2-6}
& Weathernews            & 0.359          & 0.836 & 0.825   & 0.714  \\         
\end{tabular}
\end{table*}

\subsection{Results}\label{Results}
\subsubsection{Comparison of Model Accuracy and Training Time}\label{subsection:models}
The linear model with L1 and L2 regularization (Elastic Net), LightGBM, XGBoost, Random Forest (Extremely Randomized Trees), and NN (with architecture Dense(1000, ReLU) - Dropout(ratio=0.2) - Dense(1000, ReLU) - Dropout(ratio=0.2) - Dense(100, ReLU) - Dropout(ratio=0.2) - Dense(1)) were compared by cross-validation on Fold 1, 2 and 3. Their hyperparameters were tuned using Optuna \cite{akiba2019optuna} or manual tuning.

\tabref{tab:models} presents the results of models, comparing the RMSE and the training time (for the hyperparameter-tuned models on the training data in test set). GPU (Tesla V100-SXM2-16GB) is used for NN, CPU (Intel(R) Xeon(R) CPU @ 2.00GHz) is used for other models, as well as in the following sections.

LightGBM and XGBoost demonstrated superior performance in terms of the RMSE (as these two models are quite similar). NN required a longer training duration, while Elastic Net, LightGBM, and XGBoost necessitated shorter training durations. It is reported that LightGBM is faster than XGBoost \cite{ke2017lightgbm}, although training durations fluctuated. Furthermore, the number of iterations derived from early stopping during the model tuning process was used for training on the entire training data. Hence, depending on the tuning results, the number of iterations may vary, and XGBoost could have a shorter training duration. In conclusion, we decided to use LightGBM in the following study because LightGBM slightly outperformed XGBoost in accuracy.

\begin{figure*}[t]
\centering
\begin{minipage}{0.48\hsize}
\centering
\subcaption{Temperature in 2 hours at Mt. Fuji}
\includegraphics[keepaspectratio, width=1.0\hsize]{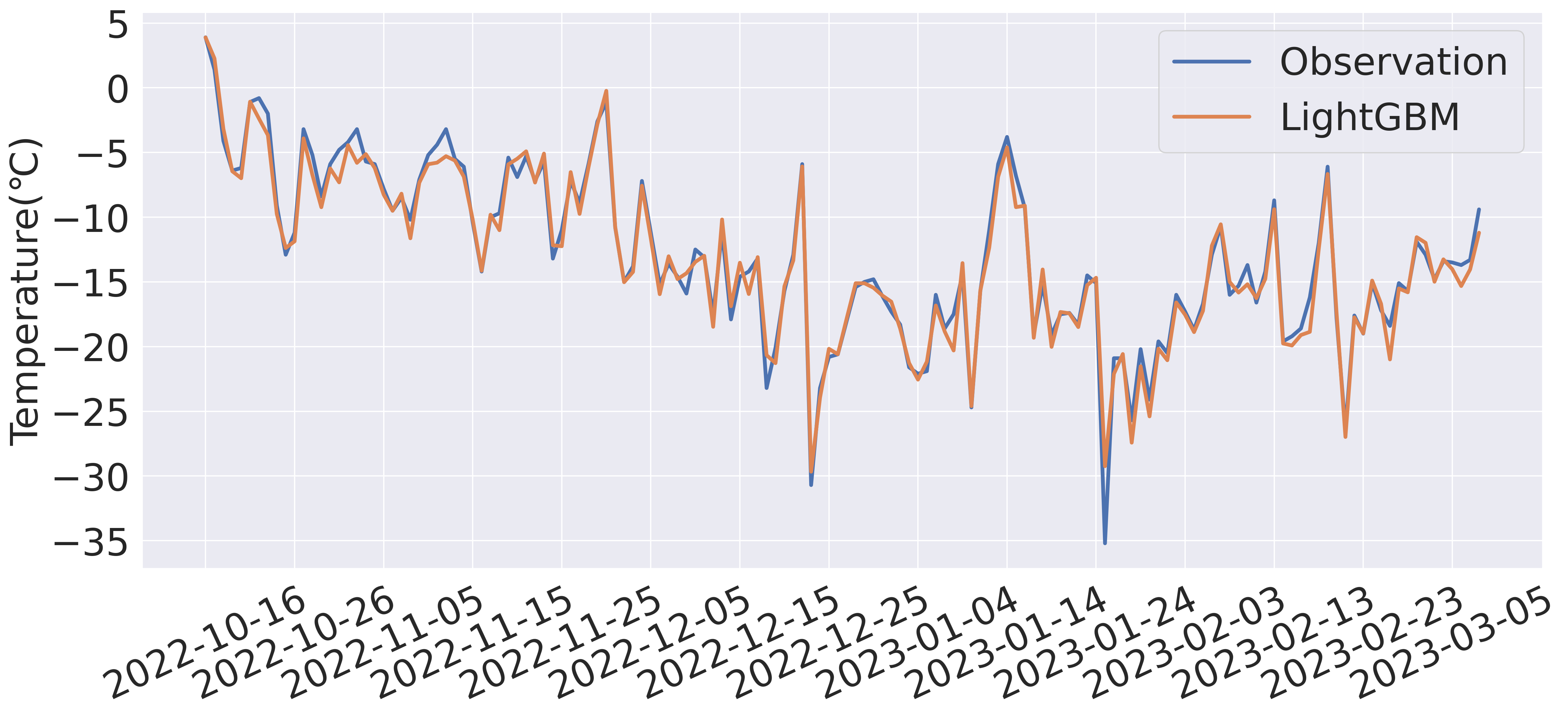}       
\end{minipage}
\begin{minipage}{0.48\hsize}
\centering
\subcaption{Precipitation in 2 hours at Hakone}
\includegraphics[keepaspectratio, width=1.0\hsize]{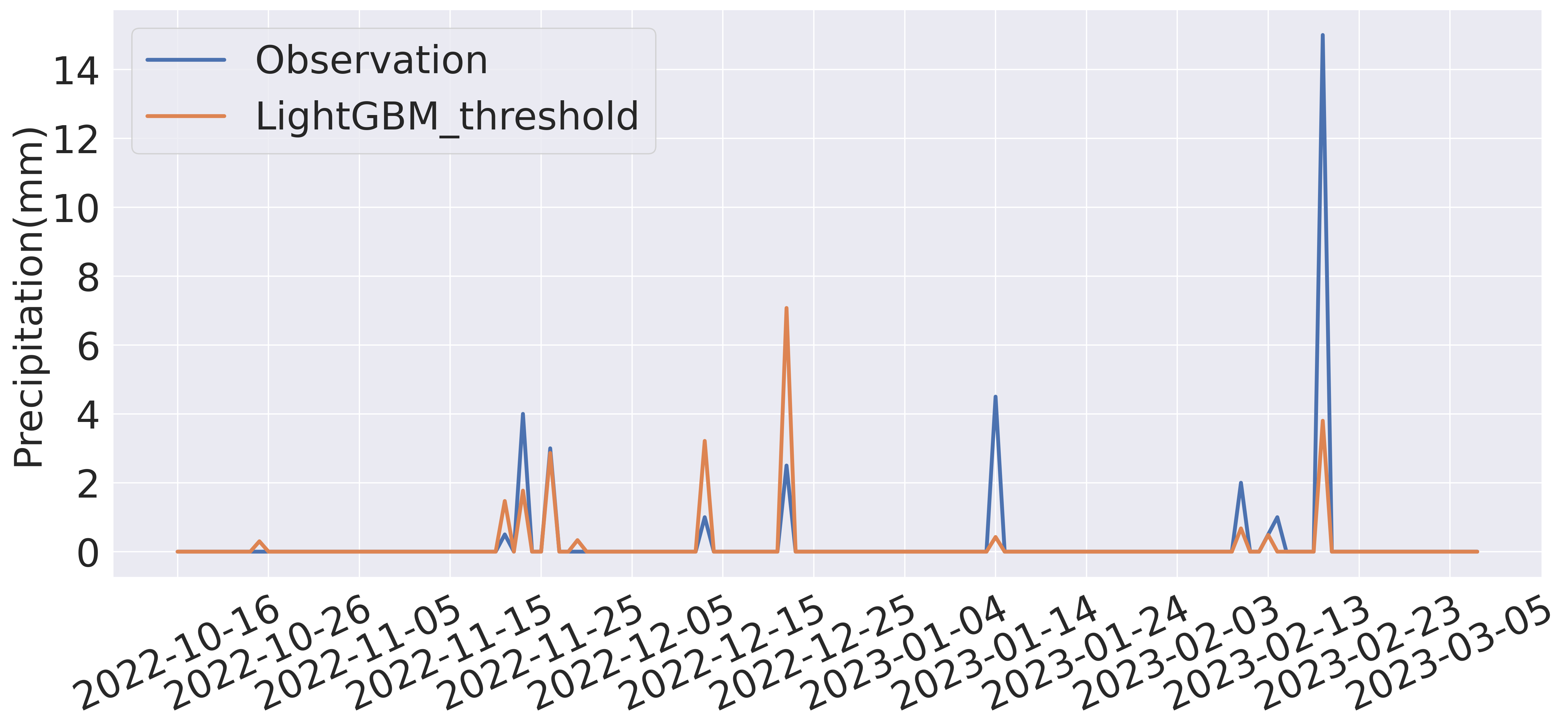}
\end{minipage}\\

\begin{minipage}{0.48\hsize}
\centering
\subcaption{Temperature in 8 hours at Mt. Fuji}
\includegraphics[keepaspectratio, width=1.0\hsize]{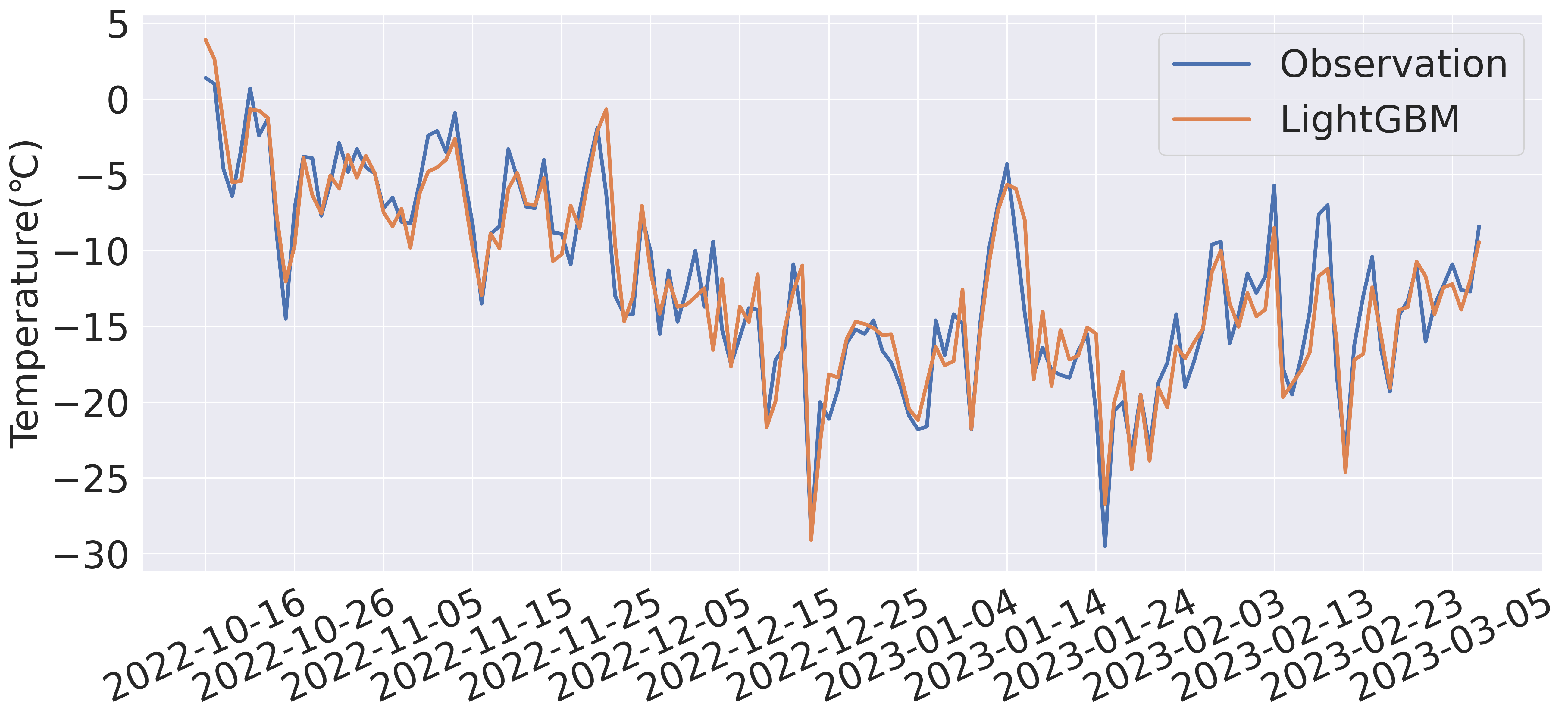}
\end{minipage}
\begin{minipage}{0.48\hsize}
\centering
\subcaption{Precipitation in 8 hours at Hakone}
\includegraphics[keepaspectratio, width=1.0\hsize]{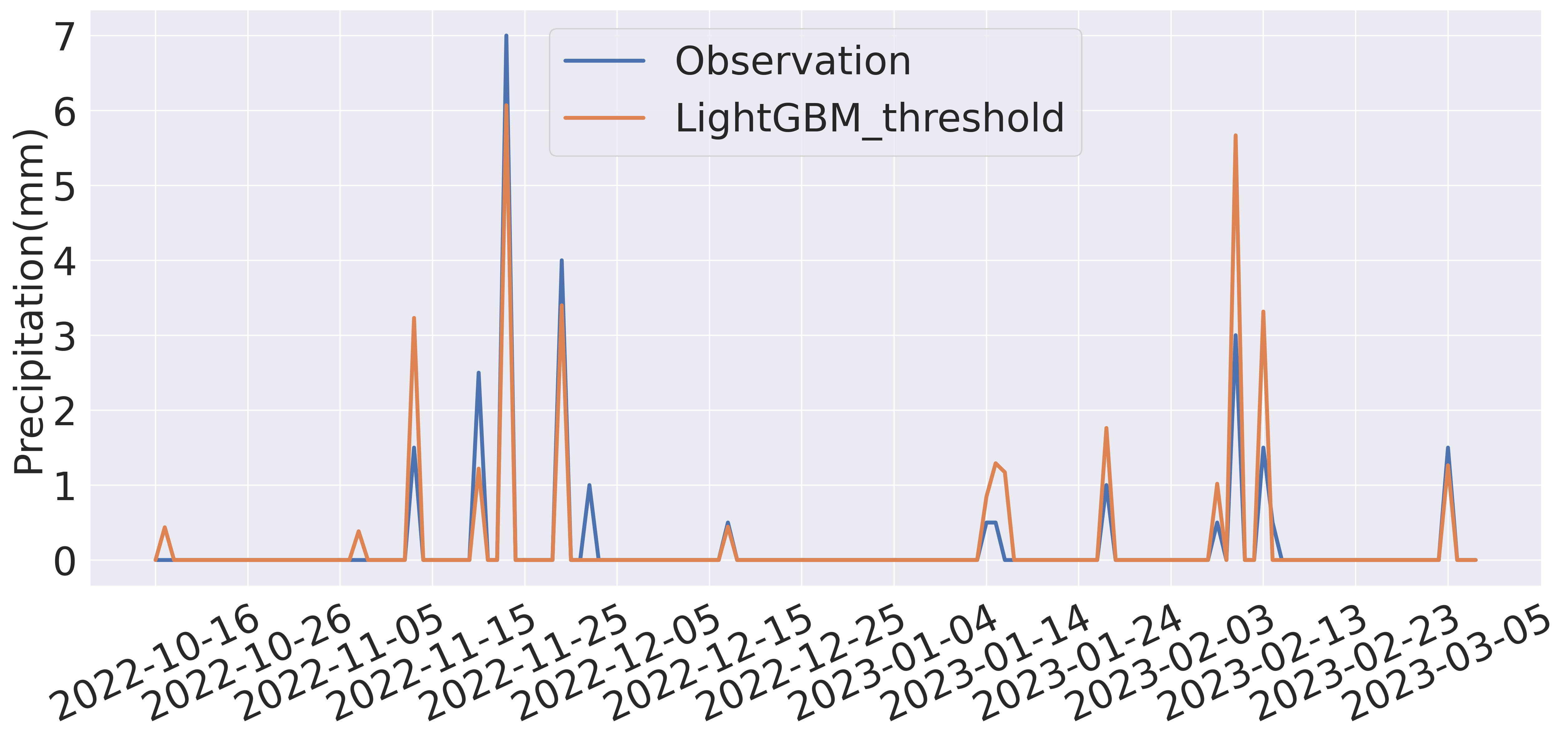}
\end{minipage}\\
\caption{Predictions in time series. LGBM (all) is used for the predictions at Mt. Fuji. LGBM binary (all) with a threshold procedure is used for the predictions at Hakone.}
\label{fig:time series}
\end{figure*}

\begin{figure*}[t]
\centering

\begin{minipage}{0.48\hsize}
\centering
\subcaption{2 hours LGBM (all)}
\includegraphics[keepaspectratio, width=0.7\hsize]{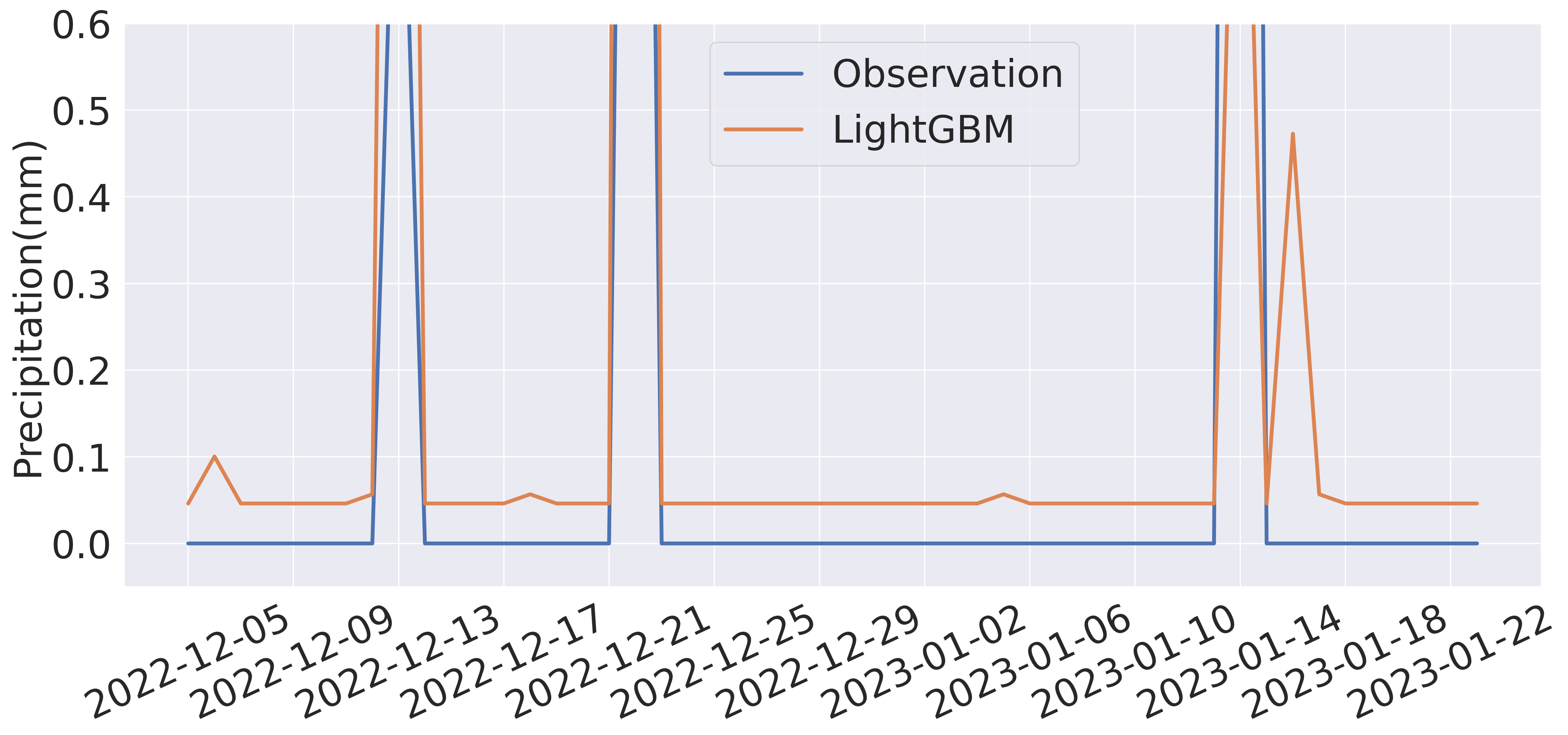}
\end{minipage}
\begin{minipage}{0.48\hsize}
\centering
\subcaption{2 hours LGBM binary (all) }
\includegraphics[keepaspectratio, width=0.7\hsize]{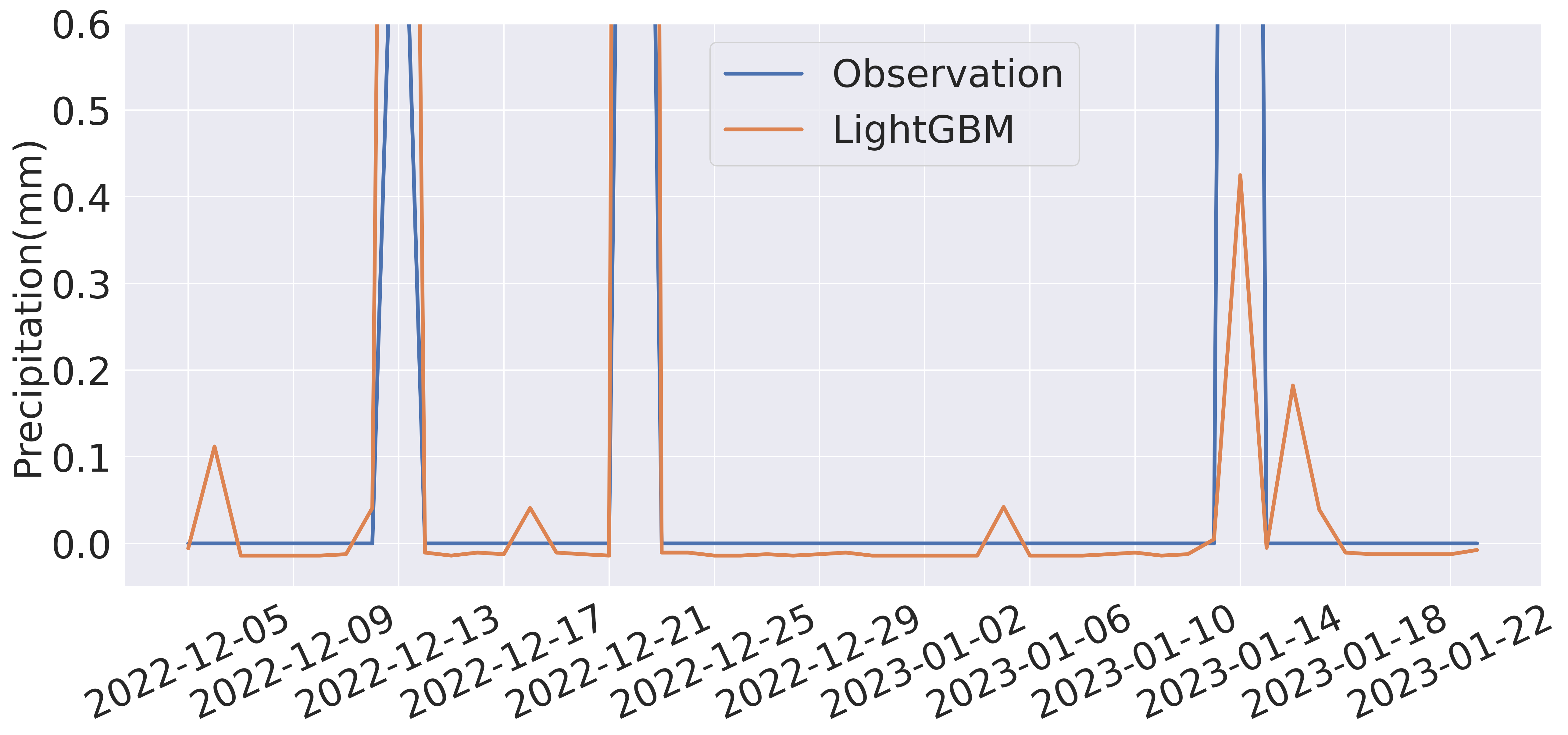}
\end{minipage}\\
\begin{minipage}{0.48\hsize}
\centering
\subcaption{8 hours LGBM (all)}
\includegraphics[keepaspectratio, width=0.7\hsize]{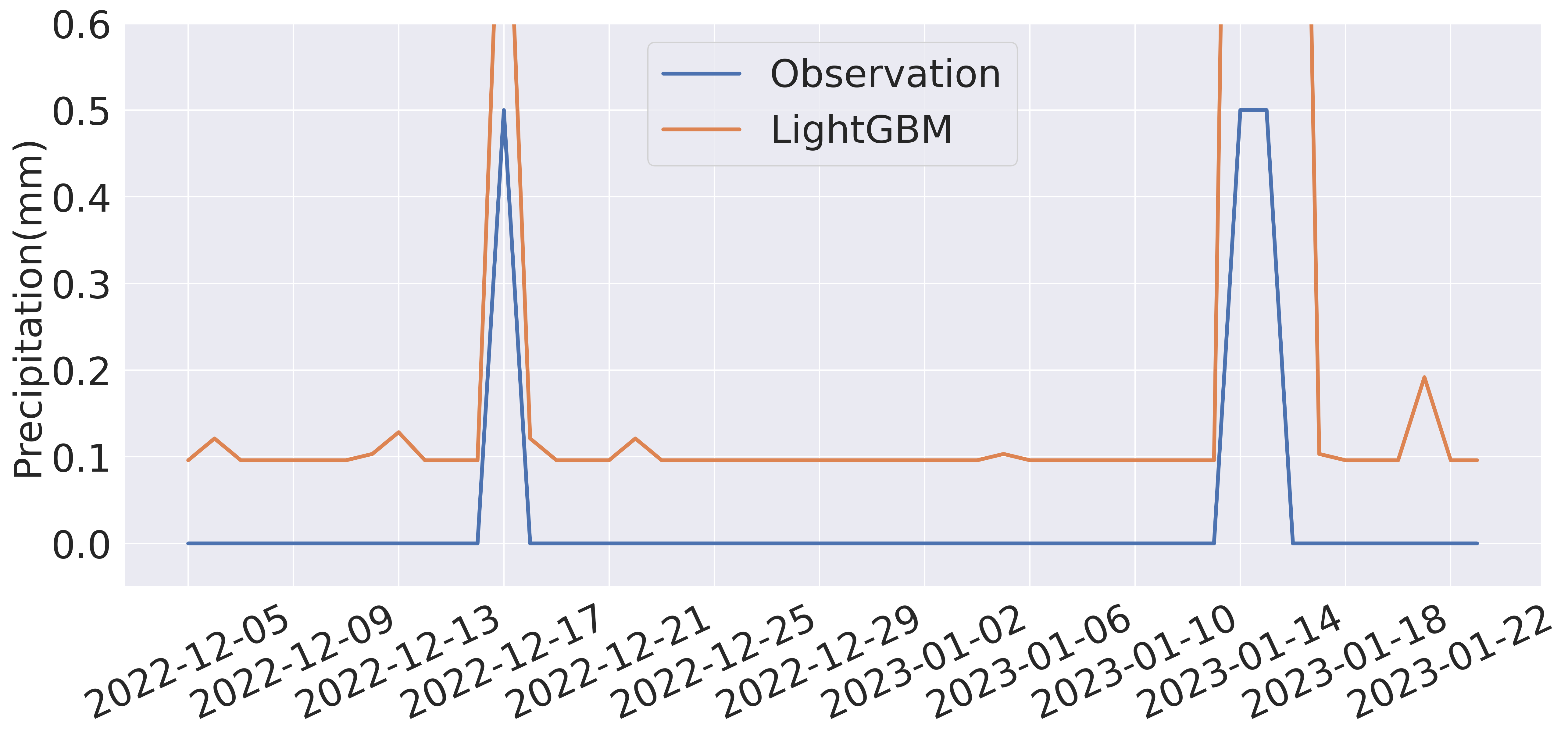}
\end{minipage}
\begin{minipage}{0.48\hsize}
\centering
\subcaption{8 hours LGBM binary (all) }
\includegraphics[keepaspectratio, width=0.7\hsize]{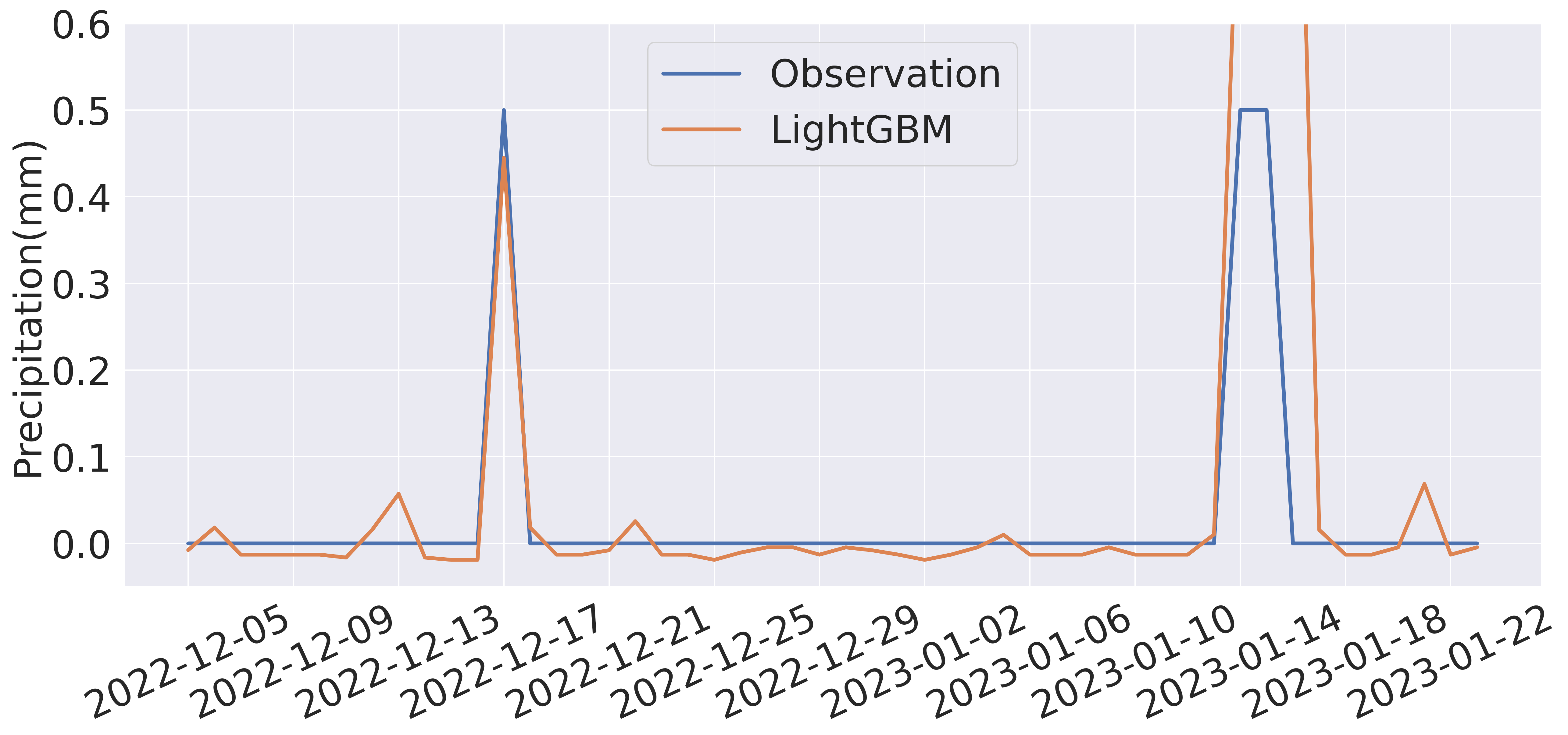}
\end{minipage}\\
\caption{Precipitation predictions in 2 hours and in 8 hours at Hakone. Threshold procedure is not applied. A bias of size about (a) 0.046 [mm] and (c) 0.096 [mm] appear in the prediction by LightGBM with MSE as a loss function, while it approaches to zero in the prediction by LightGBM using binary cross entropy in a loss function (right).}
\label{fig:bias}
\end{figure*}

\subsubsection{Feature Extraction}\label{section:Feature}
For feature extraction, we used LightGBM, which consistently demonstrated high performance in terms of accuracy and training time in the experiments described in Section \ref{subsection:models}. By performing feature extraction, we expected to reduce the inference time after training and mitigate overfitting. 

After parameter tuning, we selected 130 features for temperature prediction and 30 features for precipitation prediction based on their importance in the trained LightGBM models. These numbers were determined by experiment so that the accuracy does not drop significantly and the number of features is small. The importance of a feature is determined by counting how many times it was used for branching in the decision trees of the LightGBM algorithm. However, for the features related to time and date represented by $\sin$ and $\cos$, we adopted them regardless of their importance score.

After feature extraction, we performed parameter tuning again and created new models with selected top features. Thus, there exist LightGBM trained with all the features (LGBM (all)) and LightGBM trained with only the extracted features (LGBM (top)).

Comparing the RMSE on the test data between LGBM (all) and LGBM (top), as shown in \tabref{tab:rmse}, we observed similar or slightly worse accuracy with the extracted features. It is worth noting, however, that despite significantly reducing the number of features, there was no drastic decrease in accuracy, and a substantial reduction in training time was observed in \tabref{tab:rmse}. 

From \figref{fig:importance}, we can see that for LightGBM, past temperatures at Mt. Fuji are important for temperature prediction, while surrounding precipitations around the target time are crucial for precipitation prediction at Hakone.

\subsubsection{Loss Function using Binary Cross-Entropy}
We compared the results of training with the standard MSE loss and a linear combination of the MSE loss and the shifted binary cross entropy \eqref{binaryloss}, using the RMSE as the evaluation metric. 
We call LightGBM using this linear combination loss LGBM binary.
Regarding the validation data (\tabref{tab:binary_val}), LGBM binary (all/top) did not show much difference compared to LGBM (all/top). On the other hand, for the test data (\tabref{tab:rmse}), LGBM binary (all/top) achieved slightly better performance than LGBM (all/top), except for 2-hour forecasts.

Therefore, it can not be said that there was a significant improvement in RMSE by the introduction of binary cross-entropy (especially for validation data). However, it was observed that when the MSE was simply a loss function, there was a bias that the predicted value would be a constant non-zero value when there was no rainfall, but this bias was almost eliminated, and the predicted value approached zero by the introduction of the binary cross-entropy (\figref{fig:bias}).

\subsubsection{Comparison of LightGBM with Weather Forecasting Services}\label{compare}
We compared the predicted temperature values at Mt. Fuji and precipitation values at Hakone for 2 hours, 7 hours, 8 hours, and 9 hours ahead with the observed data, the forecasts from Weathernews, and the forecasts from Tenki to Kurasu (\tabref{tab:rmse}).
In addition, we created a model called LGBM (no future), which uses only the observed data without incorporating any forecast data and other conditions are the same as LightGBM we discussed, to examine the impact of using forecast data.

In the prediction of temperature at Mt. Fuji for the test data, LightGBM significantly outperformed the forecast services for all time points. Particularly, the accuracy of LGBM (all/top) was high for the predictions at 7, 8, and 9 hours ahead.

For the prediction of precipitation at Hakone for the test data, LGBM binary (no future + all), which used only the observed data, achieved the highest accuracy for the 2-hour ahead prediction, while LightGBM binary (all) and LightGBM binary (top) showed the highest accuracy for the predictions at 8 and 9 hours ahead, respectively.

Furthermore, it should be noted that the accuracy for the predictions at 7, 8, 9 hours ahead is higher than that for the 2-hour ahead prediction for precipitation at Hakone. This discrepancy is attributed to the limited size of the test data (144 data), which made it more susceptible to the influence of outliers such as 15.0 [mm] at 9 a.m. on 2023-02-19 (\figref{fig:time series}).

We also compare LightGBM with Weather Forecasting Services using the correlation coefficient as an evaluation metric  (\tabref{tab:cc}).  
The results were similar to the RMSE case, but differences between temperature predictions by ``Tenki to Kurasu'' and by LightGBM become relatively smaller. This would imply that the linear interpolation for ``Tenki to Kurasu'' we adopted to adjust the elevation (as the caption of \tabref{tab:RMSETest}) remained biased.

\section{Conclusion}
The experiments in this paper aimed to predict weather in mountainous areas by interpolating forecasts for the surrounding regions. 
The GBDT model outperformed in terms of both accuracy and training speed. On the other hand, feature extraction was effective only in terms of training time. In predicting precipitation, the use of a linear sum of the MSE and the shifted binary cross entropy as the loss function improved the bias when there is no rainfall.

Our method outperformed some existing weather services in predicting the temperature at Mt. Fuji. However, no significant improvement was observed in predicting precipitation at Hakone 7, 8, and 9 hours ahead of time. This could be attributed to the comparable accuracy of the precipitation forecast for Hakone and the surrounding areas. In addition, it should be noted that our models are trained on a small amount of observed data from only eight sites and just over three years.

 
In this paper, observed data were used instead of forecast data for the surrounding areas in the future during training. 
One reason for doing so is that historical forecast data is difficult to obtain from publicly available sources.
Another concern with training on forecast data is that modification of forecast models may affect the training of our model.
Nevertheless, our model was reasonably accurate, which supports the assumption that the forecasting data in surrounding area is free of biases, such as constant errors.
 
Since NWP or its guided version are grid data, they can be used as input to create versatile models that do not depend on the location or weather element to be predicted. However, since they are grid data, the dimension of the data is high, and even when forecasting a single weather element at a single location, the appropriate machine learning model and input features will be different from those in this paper. In addition, the bias is expected to be greater than for forecast data fitted to observation points. Therefore, it is not certain that a simple application of the method in this paper will produce highly accurate prediction.

\subsection*{Acknowledgments}
KI was supported by the WISE Program for the Development of AI Professionals in the Marine Industry at TUMSAT. TT was supported by the Japan Society for the Promotion of Science, Grand-in-Aid (C) (22K03383).

\appendix
\section{Binary Cross-Entropy Loss}
\label{A1}
Let $y$ and $\hat{y}$ be the target value (ground truth)  
and predicted value of precipitation respectively, and let $z$ and $\hat{z}$ be 
a binarized value with a threshold of $0.25$ for $y$ and
a value of a shifted sigmoid function at $\hat{y}$ respectively as
\begin{equation}
  z =
  \begin{cases}
    1 & \text{if $y\geq 0.25$} \\
    0 & \text{if $y< 0.25$}
  \end{cases}
\end{equation}
\begin{equation}
\hat{z} = \frac{1}{1+e^{-a(\hat{y}-0.25)}}, 
\end{equation}
where $a$ is a constant. 
We define a linear combination of the MSE loss and the shifted binary cross entropy 
as \eqref{binaryloss}, where the MSE is\footnote{For regression problems, LightGBM uses the MSE with half as the loss function as \eqref{eqmse} by default.} 
\begin{equation}
\label{eqmse}
{\rm MSE} = \frac{1}{2N} \sum_{i=1}^{N} (y_i - \hat{y})^2
\end{equation}
and 
$L_{{\rm binary}}$ is a binary cross-entropy of $\hat{z}$ for $z$: 
\begin{equation}
L_{{\rm binary}}= \frac{1}{N}  \left(- z \log \hat{z} - (1-z) \log (1-\hat{z})\right),
\end{equation}
where $N$ is the number of samples in the dataset.

\begin{figure}[t]
\centering
\begin{minipage}{0.7\hsize}
\centering
\subcaption{In the case of $y=0.0$}
\includegraphics[keepaspectratio, width=0.9\hsize]{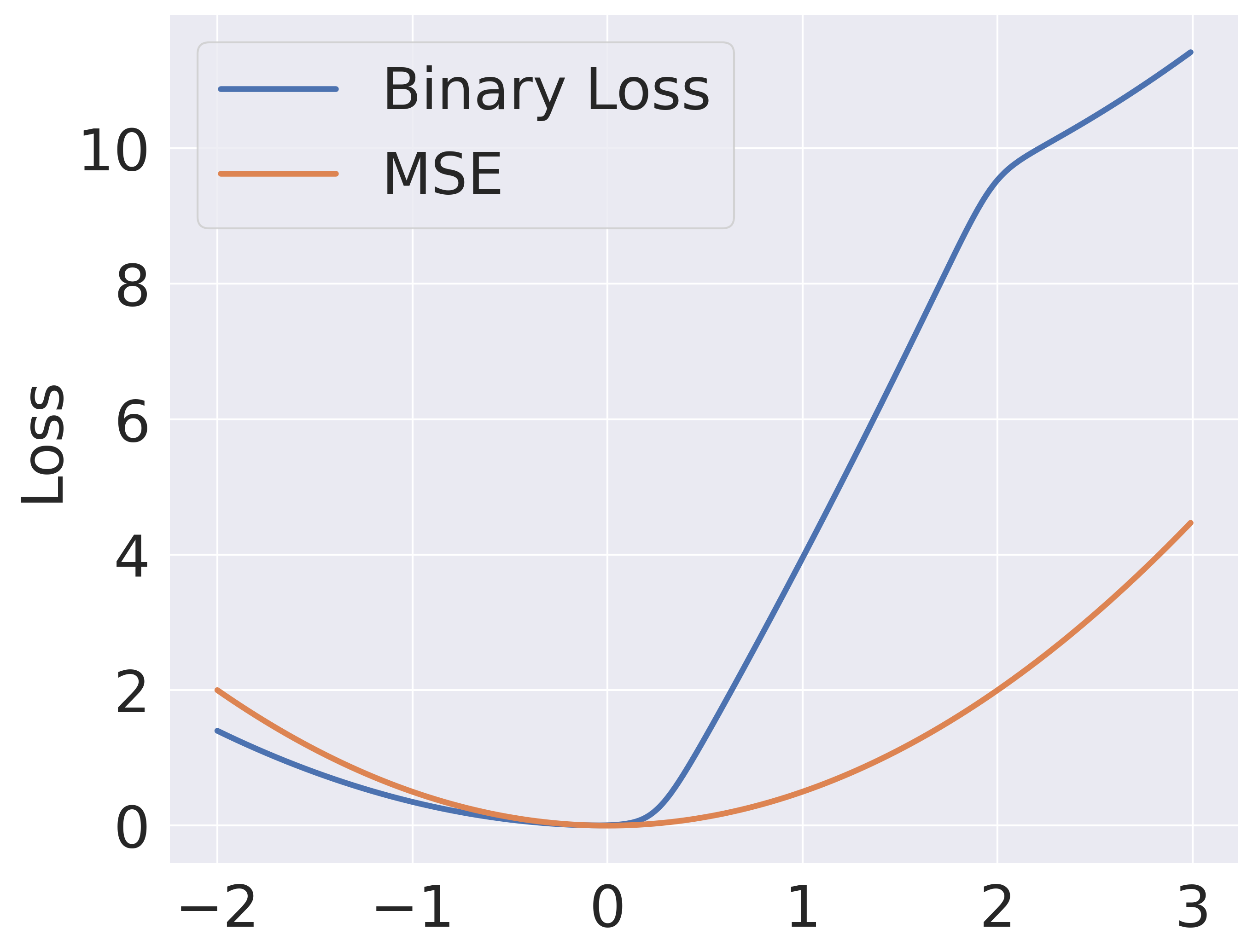}
\end{minipage}\\
\begin{minipage}{0.7\hsize}
\centering
\subcaption{In the case of $y=0.5$}
\includegraphics[keepaspectratio, width=0.9\hsize]{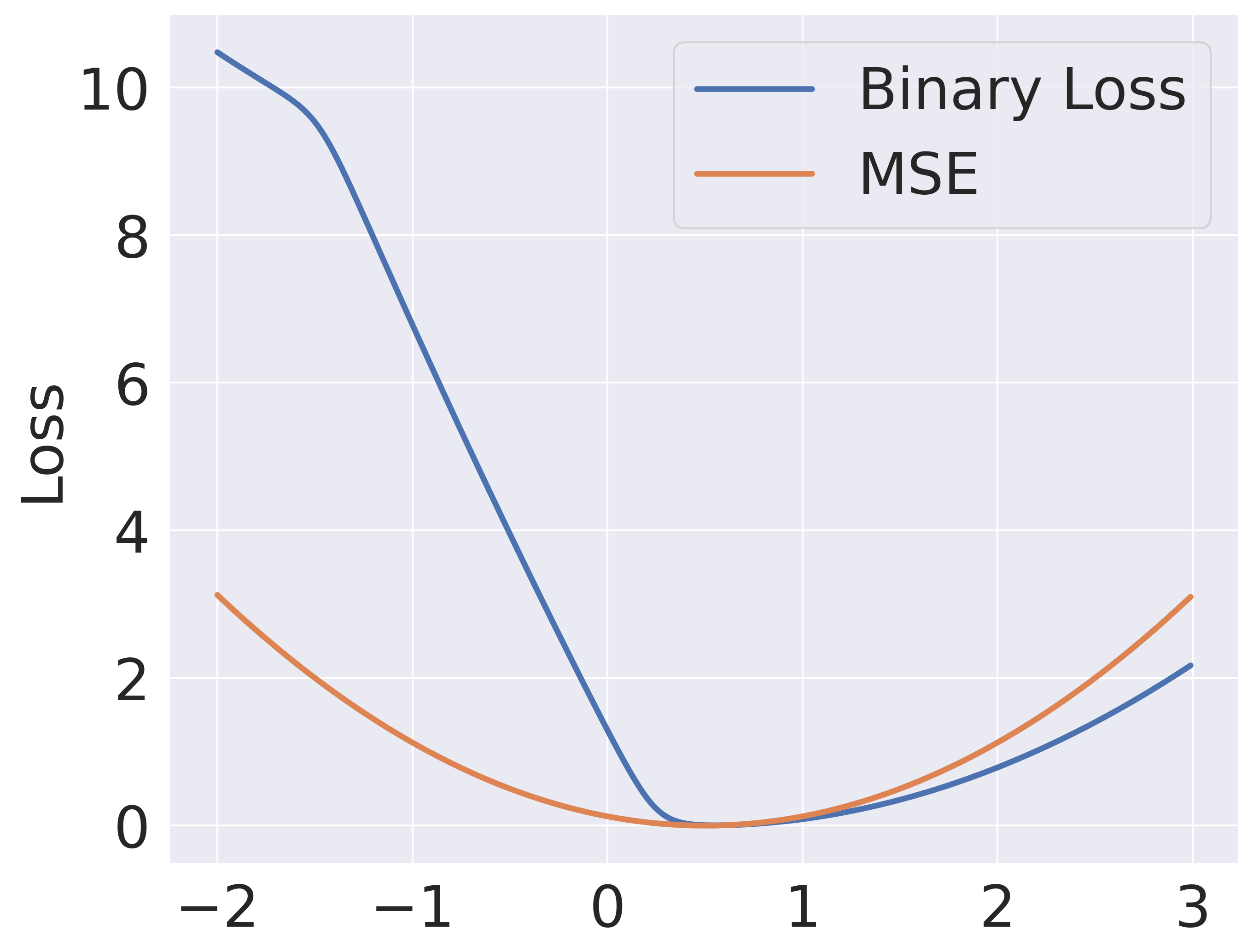}
\end{minipage}
\caption{MSE and $L_{{\rm binary}}$ (Binary Loss) for $y=0.0$ and $y=0.5$ when changing $\hat{y}$. Horizontal axis: $\hat{y}$}
\label{fig:binaryloss}
\end{figure}

\begin{figure}[t]
\centering
\begin{minipage}{0.75\hsize}
\centering
\subcaption{Precipitation in 2 hours at Hakone}
\includegraphics[keepaspectratio, width=0.9\hsize]{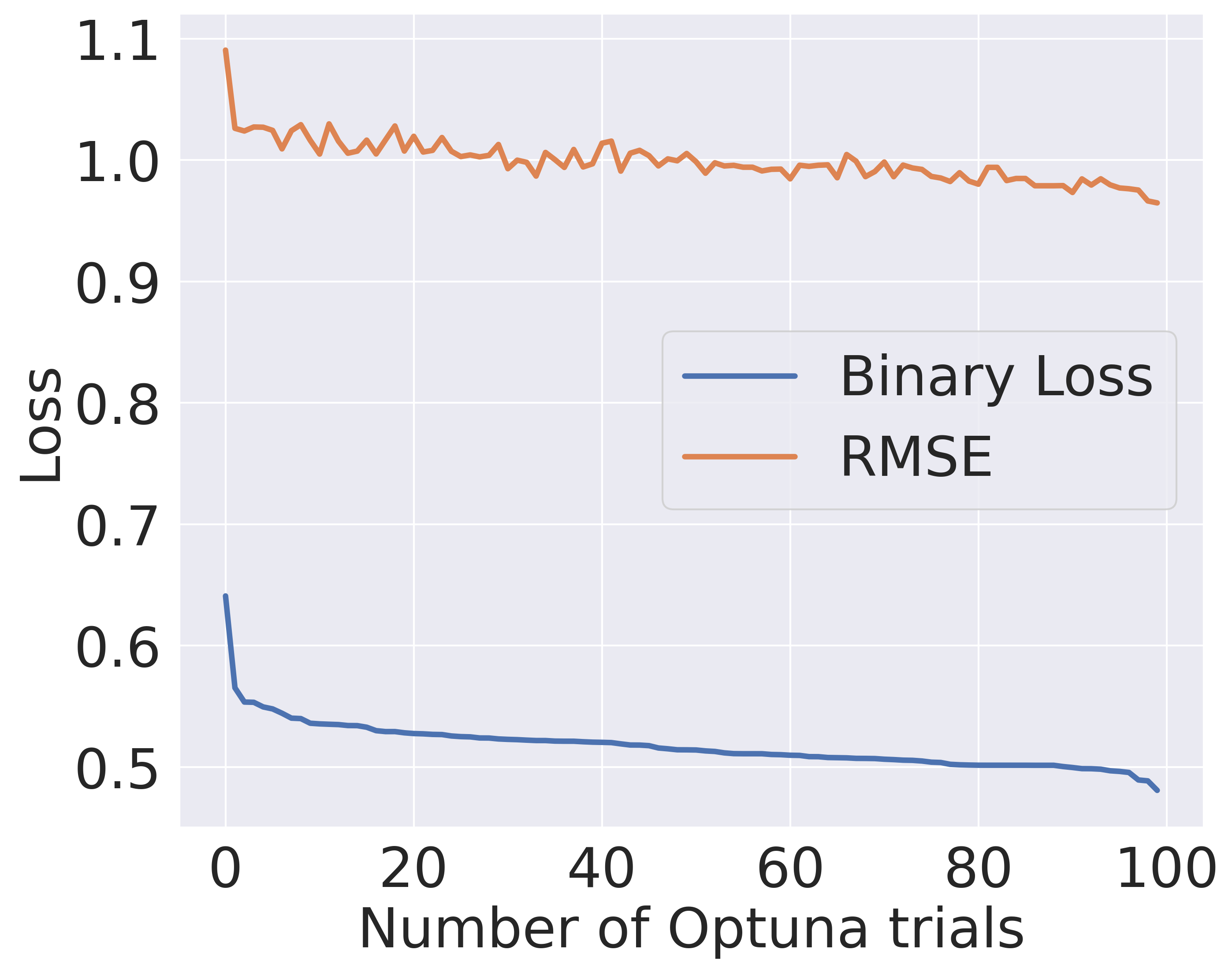}
\end{minipage}\\
\begin{minipage}{0.75\hsize}
\centering
\subcaption{Precipitation in 8 hours at Hakone}
\includegraphics[keepaspectratio, width=0.9\hsize]{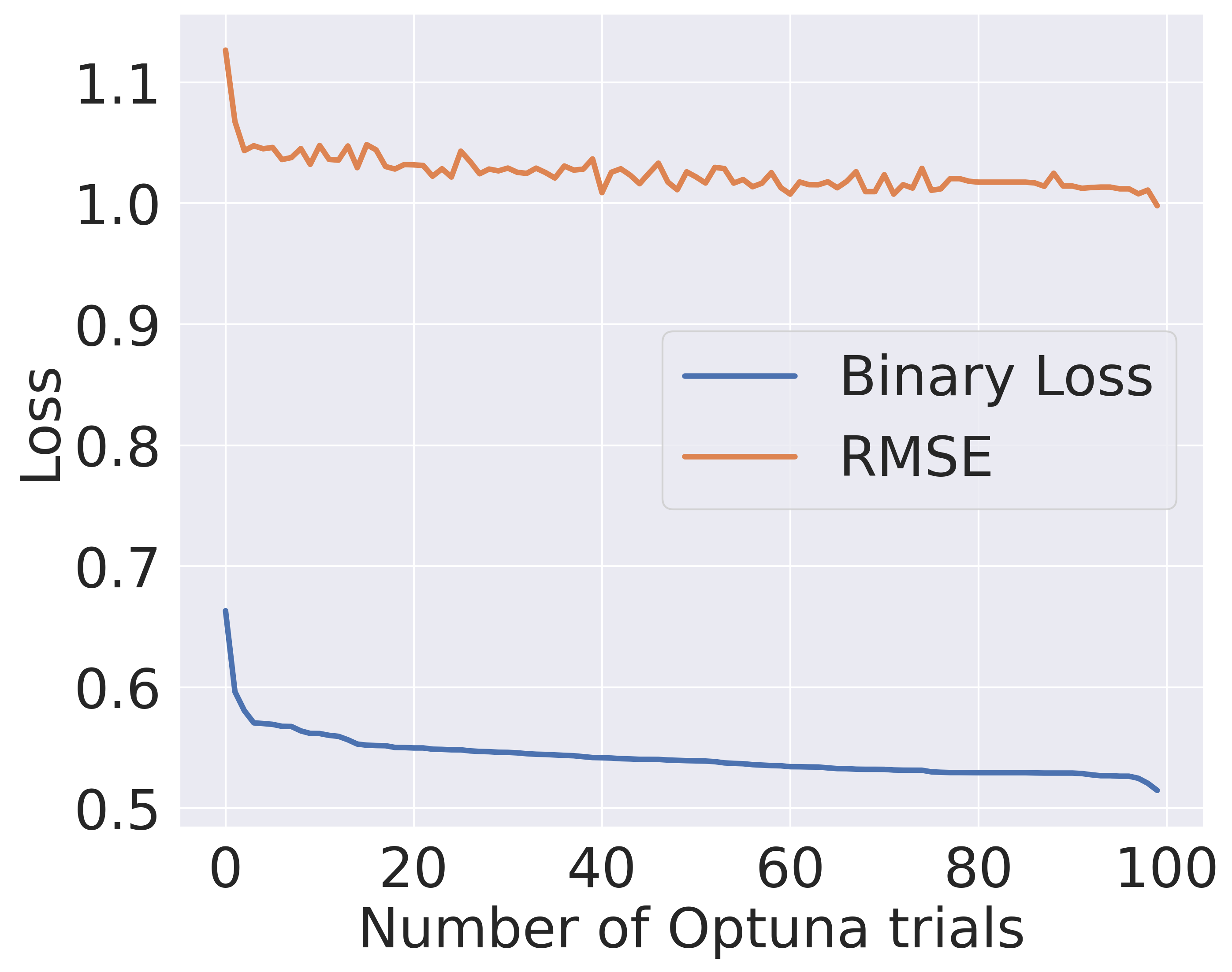}
\end{minipage}
\caption{Comparison of the RMSE and the $L_{{\rm binary}}$ (Binary Loss) losses during tuning for all features by Optuna. Note that the order of trials is sorted by $L_{{\rm binary}}$.}
\label{fig:losscomparison}
\end{figure}

Specifically, we set as $a$ = 16, $\alpha$ = 0.7 in this work. This value was determined through experimentation to optimize the influence of binary cross-entropy while preserving acceptable accuracy on the validation data. Note that the graph of the loss function, as shown in \figref{fig:binaryloss}, indicates that $L_{{\rm binary}}$ emphasizes whether $\hat{y}$ is greater than 0.25 compared to the MSE. When using $L_{{\rm binary}}$, we also used it as an objective for the tuning by Optuna. The relationship between $L_{{\rm binary}}$ and the RMSE during the tuning process is shown in \figref{fig:losscomparison}. The values of $L_{{\rm binary}}$ and the RMSE are generally correlated, showing that as $L_{{\rm binary}}$ decreases, the RMSE tends to decrease as well.
\end{document}